\documentclass[twocolumn,showpacs,preprintnumbers,amsmath,amssymb,aps]{revtex4-1}

\usepackage{graphicx}
\usepackage{dcolumn}
\usepackage{bm}
\usepackage{color}

\usepackage[normalem]{ulem}
\usepackage{amssymb,amsmath}

\begin{document}
\title{Experimental Investigation of Plastic Deformations Before Granular Avalanche}

\author{Axelle Amon} \email{axelle.amon@univ-rennes1.fr} \author{Roman
  Bertoni} \author{J\'er\^ome Crassous}
\affiliation{Institut de Physique de Rennes, UMR UR1-CNRS 6251,
  Universit\'e de Rennes 1, Campus de Beaulieu, F-35042 RENNES cedex,
  France}

\date{Received: date / Revised version: date}

\begin{abstract}
We present an experimental study of the deformation inside a granular
material that is progressively tilted. We investigate the deformation
before the avalanche with a spatially resolved Diffusive Wave
Spectroscopy setup. At the beginning of the inclination process, we
first observe localized and isolated events in the bulk, with a
density which decreases with the depth. As the angle of inclination
increases, series of micro-failures occur periodically in the bulk,
and finally a granular avalanche takes place. The micro-failures are
observed only when the tilt angles are larger than a threshold
  angle much smaller than the granular avalanche angle.  We have
characterized the density of reorganizations and the localization of
micro-failures. We have also explored the effect of the nature of the
grains, the relative humidity conditions and the packing fraction of
the sample.  We discuss those observations in the framework of the
plasticity of granular matter. Micro-failures may then be viewed as
the result of the accumulation of numerous plastic events.
\end{abstract}


\maketitle


\section{Introduction}

The onset of avalanches in granular materials is a well-documented
area of research motivated by evident applications in the field of
risk prevention. A typical granular experiment to study such
phenomenon consists in the slow inclination of a sand filled box until
the destabilization of the sand
pile~\cite{jaeger89,bretz92,nerone03}. A lot of studies have been
devoted to the angles at which this avalanche initiates and
stops~\cite{nagel92,fischer08}. The critical angle, or avalanche
angle, is then defined as the angle at which a large flow of grains
occurs, with the surface of the heap stabilizing at a smaller angle,
the angle of repose. Usually studies of stability of granular heaps
focus on the succession of avalanches~\cite{jaeger89}, either in
rotating drums or by using a local perturbation of a heap. In those
two cases, the heap has been built by successive avalanches, giving a
specific internal structure to the bulk. The time intervals between
successive events and the size of the avalanches is then discussed,
one of the points being to know if those distributions are or are not
power-law distributions as predicted for a critical
phenomena~\cite{bak87}. In the case of a granular pile without initial inner structure progressively inclined in the gravitational field, specific types of rearrangements taking place before the first avalanche have been identified. Experimentally, two kinds of
movement in the heap previous to its failure have been detected:
\emph{small rearrangements}, implying only a few grains, and periodic
\emph{precursors} of large
amplitude~\cite{nerone03,zaitsev08,gibiat09,kiesgen12}. Most of the
studies reporting this phenomenology are based on the observation of
the free surface of the pile, but recent works give clues that these
precursors are bulk phenomena~\cite{zaitsev08,gibiat09}. To our
knowledge, these regular events have never been reproduced numerically
or explained theoretically, although preavalanche instabilities
showing an intermittent evolution have been studied
numerically~\cite{staron02,staron06}.

The periodicity of the precursors are reminiscent of the
\emph{stick-slip} phenomenology, which is usually invoked as an
explanation of the observed regularity. To our knowledge, a
  complete interpretation in that framework, including the prediction
  of the starting angle of the process, which is well beyond the angle
  of friction of the material, or of the selected period, has never
  been made. \emph{Stick-slip} behavior emerges when a frictional
system is submitted to an increasing load at a small enough
rate~\cite{duran}. It originates from the difference between the
static friction and the dynamic friction. In the simplest model, a
traction is exerted on a frictional slider by a spring. Elastic energy
is first stored in the spring while the slider is stuck until the
tangential force is large enough to overcome the static friction. The
slider then enters an oscillating response that leads it to stick
again when the velocity vanishes. A model inspired by this principle
has been used to explain the succession of avalanches in a rotating
drum~\cite{duran}. In the same spirit, arrays of sliders connected by
springs (Burridge-Knopoff model)~\cite{carlson94} are used to model
seismic processes. Fault gouges are often modeled in the geology
literature by a granular material sheared between two planes and the
\emph{stick-slip} behavior of such systems has been extensively
studied in an established regular regime of shearing of the granular
layer~\cite{nasuno97,nasuno98}. The existence of periodicity in
seismology is still under discussion and recent models exhibit the
coexistence of quasiperiodicity and criticality~\cite{ramos06}.

It is striking to observe that a very similar phenomenology of regular
precursors of rupture before the establishment of the regular
\emph{stick-slip} response can be found in the literature in two other
configurations closely linked to the problem of failure in granular
materials: the onset of frictional motion and the mechanical response
of metallic glasses. In the case of frictional motion, regular
precursors are observed before the onset of
sliding~\cite{rubinstein07,rubinstein09,ben-David10,maegawa10}. The
present interpretation relies on the inhomogeneity of the spatial
repartition of the normal and shear stresses at the interface leading
to the possibility of reaching the Coulomb criteria in some localized
areas before that the whole system is
destabilized~\cite{schreibert10,tromborg11}. The links with the models
of earthquakes described in the previous paragraph have been
made~\cite{braun09,rubinstein11}. In the second system, the case of
the mechanical response of metallic glasses, the observed
\emph{stick-slip} phenomenology is called \emph{serration}. The
mechanical response of such amorphous glasses is very close to the
response of granular material and displays similar features: creep,
shear-banding and precursors to the rupture, all of which seem to be
generic to amorphous materials. Comparing the loading curves obtained
for metallic glasses (Figure 1(a) in~\cite{klaumunzer11}), granular
materials (Figure 2(b) in~\cite{nguyen11}) and onset of friction
(Figure 1(a) in~\cite{rubinstein07}) a striking similarity is
observed. Each of these three curves displays small regular stress
drops during the otherwise quasi-elastic loading until the failure (or
sliding) of the material. Those drops begin to appear well before the
rupture of the material or the onset of motion of the slider. After
the yield, the system enters a regular stick-slip motion. In the case
of granular materials, it has been shown that the value of the stress
at the first micro-rupture event is linked to the stress at which the
global yield will take place~\cite{nguyen11}: independent of the
packing fraction of the system, the ratio between those two stresses
is constant. Using a method of detection identical to the one that
will be used in this paper, it has been shown that the drops in the
loading curve are linked to the appearance of a shear band in the
system~\cite{amon12}. During the loading of the system, ``spots'' of
localized strong deformation are also observed. As it will be shown,
these localized spots tends to appear at a higher rate, and to
cluster, just before a shear band event occurs~\cite{amon12}.

Consequently, a natural question emerges regarding the understanding
of the nature of those precursor events and their potential link to
the small rearrangements. With this aim in mind, in this article, we
study experimentally the response of granular material to a
progressive, quasi-static, inclination. We will focus on the
description of the behavior of the heap before the first
avalanche. For the study of the rearrangements and of the precursors,
we use an original method of measurement of small deformation based on
Diffusive Wave Spectroscopy~\cite{erpelding08,erpelding10}. This
method gives access to the phenomenology of the rearrangements and the
precursors at the side of the sample, and consequently throughout the
depth of the pile, giving hints of what happens in the bulk of the
system.

The structure of the article is as follows: in part \S\ref{sec2} we
will describe the mechanical setup (\S\ref{sec2.1}), the
interferometric method of detection of the rearrangements and
deformations (\S\ref{sec2.2}), and the granular materials and the
experimental protocols of preparation of the samples
(\S\ref{sec2.3}). In the third section we will present the observed
experimental behavior. After an overview of the experimental
observations (\S\ref{sec3.1}), we will characterize the periodic
\emph{precursor} events (\S\ref{sec3.2}) and then the \emph{small
  rearrangements} phenomenology (\S\ref{sec3.3}). We also discuss the
influence of several parameters (shape of the grains, humidity and
packing fraction of the sample) on the observed phenomenology
(\S\ref{sec3.4}). The results are discussed in \S\ref{sec4}. In
\S\ref{sec4.1} we compare the observations with the classical
Mohr-Coulomb theory of failure. In \S\ref{sec4.2}, we interpret these
events in the framework of plasticity of granular materials, and
coupling between plastic events.


\section{Experimental set-up}\label{sec2}
\subsection{Mechanical setup}\label{sec2.1}

\begin{figure}[htbp]
  \centering
  \includegraphics[width=\columnwidth]{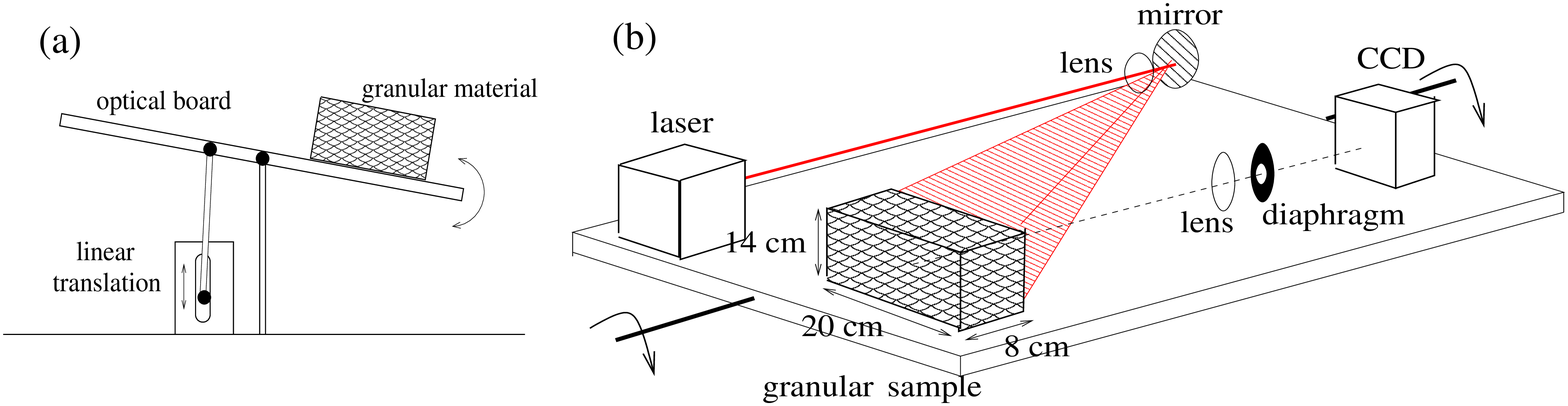}
  \caption{(a) Schematic of the crankshaft system with the tilting
    optical board. (b) Schematic of the setup clamped to the tilting
    optical table. The laser source is expanded in order to illuminate
    the side of the sample. The backscattered light is collected on a
    CCD camera. A lens images the side of the sample on the CCD and a
    diaphragm controls the size of the speckles on the camera.}
  \label{setup}
\end{figure}

The mechanical set-up consists of an optical board that can be
inclined continuously by a crankshaft system (see
Fig.~\ref{setup}(a)). The laser source that illuminates the sample,
the optical detection setup and the granular container are all fixed
on the board so that the whole set-up rotates with the granular pile
(see Fig.~\ref{setup}). The translation of the crankshaft is carried
out by a motorised linear translation stage (Micro Controle
UT100-50PP) controlled electronically (Micro Controle ITL09). This
system ensures a low level of vibration at a small rotation rate. For
translations between 25 $\mu$m/s and 100 $\mu$m/s, the tilt rates
range between 0.02$^{\circ}$/s and 0.08$^{\circ}$/s. We will show in
the following that all the experiments can be considered to be done in
a quasi-static limit.

The granular sample is contained in a rectangular box of dimensions 20
cm $\times$ 14 cm $\times$ 8 cm, the lateral sides of which are glass
windows. One lateral side is illuminated by a laser (Melles Griot, 15
mW, 633 nm). The beam of the laser is expanded by a lens in order to
illuminate the entire surface of one side of the box. A lens images
that side of the sample on a CCD camera (Prosilica GC 2450,
2448$\times$2050 pixels, cell size 3.45$\times$3.45 $\mu$m), with a
magnification $\gamma =0.09$. Because of the disorder of the granular
media, the backscattered light collected has a speckle pattern in
intensity. A diaphragm allows us to adjust the size of these speckles
on the CCD camera (see next section for the optical method). The
camera is interfaced with a computer and a Labview program ensures the
acquisition of successive images at a frame rate that has been chosen
between 1~fps and 3~fps in the experiments shown here.

\subsection{Optical method}\label{sec2.2}
The measurement of deformation and rearrangements in the granular
material is carried out using an interferometric technique based on
Diffusive Wave Spectroscopy (DWS). The method has been extensively
described in~\cite{erpelding08,erpelding10}. The treatment of raw
experimental images is based on the correlation of backscattered
intensity between two successive images following a multi-speckle
procedure. The link with the corresponding deformations that have
occurred inside the sample is made using a
model~\cite{crassous07,erpelding08}.

The schematic of the optical part of the setup is shown in
Figure~\ref{setup}(b). The beam of a laser source is expanded to
illuminate the granular sample and the side of the sample is imaged on
a camera with a lens. Because of the disordered structure of the
scattering material, light rays inside the media follow random
paths. The backscattered rays explore a typical zone of length $l^*$
in the sample, called the transport mean free path of
light. For glass spheres of diameter $d$, $l^* \simeq 3.3
  d$~\cite{crassous07}. As the incident light is coherent, those
backscattered rays interfere and give a speckled structure to the
collected intensity. The size of the coherence areas of the speckle in
the image plane is controlled with a diaphragm. When the sample is
deformed, the scatterers are displaced and the light paths modified so
that the interference pattern changes. These changes can be quantified
by calculating the correlation of intensities between two states.

In a backscattering configuration, a light ray typically
  explores a volume of characteristic size $(l^*)^3$ inside the
  scattering material. By using domains of size $(\gamma l^*)^2$ in
  the images for the ensemble average, we can then obtain a local
  evaluation of the correlation function with the optimal spatial
  resolution. Correlation maps are thus obtained, giving insights into
  the processes that have taken place in the sample during the
  loading.

The intensity correlation between two successive images, labelled 1
and 2, is calculated using the relation:
$$ G_I = \frac{\langle I_1 I_2 \rangle - \langle I_1 \rangle \langle
  I_2 \rangle}{\sqrt{\langle I_1^2 \rangle - \langle I_1 \rangle^2}
  \sqrt{\langle I_2^2 \rangle - \langle I_2 \rangle^2}},$$ where $I_1$
and $I_2$ are the scattered intensities on images 1 and 2. The
electronic noise of the camera and possible stray light are taken into
account with a procedure explained in~\cite{djaoui2005}. The
  ensemble averages $\langle \cdot \rangle$ are calculated over square
  sub-areas of the images of size 16$\times$16 pixels, which typically
  contain about forty coherence areas. From initial images of size
  2448$\times$2050 pixels we then obtain maps of 153$\times$128
  pixels. With the lens magnification chosen here, the spatial
  resolution of the final map is about 608 $\mu$m.

The map of correlations are calculated between successive images,
giving information about the incremental decorrelations that have
taken place in the system for an angle increment. When the
  tilt rate was modified, the frame rate of the camera was chosen to
  ensure that the angle increment between two successive images stayed
  between 0.06$^{\circ}$/slice and 0.08$^{\circ}$/slice. This allowed
  for consistency and ease of comparison between the different
  experiments.

In the case of linear elastic samples~\cite{erpelding08}, the
  deformation of the material corresponding to the measured
  decorrelation can be estimated using a model. The extension of that
  model in the case of granular materials has been discussed
  in~\cite{erpelding10}. Considering that the deformation between the
two correlated images is affine at the scale of $l^*$ in the material,
that the light is strongly scattered in the sample, and that we
observe the backscattered light, the expected dependence of the
intensity correlation function on the strain tensor is as follows:
\begin{equation}
G_I \simeq \exp (-c\sqrt{f(\mathbf{U})}),\label{GI}
\end{equation}
where $f(\mathbf{U})$ is a quadratic function of the strain tensor
$\mathbf{U}$, (see~\cite{erpelding08,erpelding10} for details of that
function), corresponding to the deformation increment that has taken
place between the two states associated with the images used for the
calculation of the
correlation~\cite{erpelding08,erpelding10}. Following~\cite{erpelding10},
the constant $c$ has been taken as $c=2 \eta \frac{2\pi}{\lambda} l^*
\sqrt{\frac{2}{5}}$, with $\eta \simeq 2$ a constant depending of the
optical setup and $\lambda$ the wavelength of the laser. The typical
deformations probed by this method for values of the transport length
similar to the one of the glass beads used here, $l^* \simeq
990~\mu$m, are between $10^{-6}$ and
$10^{-4}$~\cite{erpelding10,amon12}. For sand, we measured $l^* \simeq
660~\mu m \simeq 2d$. This lower value of $l^*/d$ compared to glass
beads may be understood by noticing that the surface of sand grains
should itself diffuse light. However, since the ratio $l^*/d$ is not
very different for glass beads and sand, we expect the magnitude of
probed deformations to be very similar.

\subsection{Granular samples and preparation}\label{sec2.3}
\begin{figure}[htbp]
  \centering
  \includegraphics[width=\columnwidth]{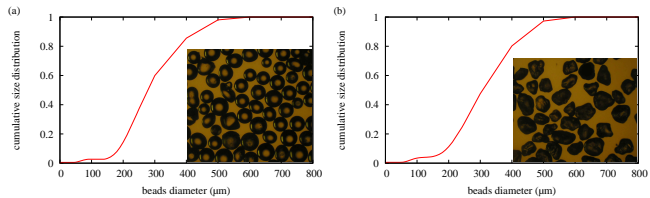}
  \caption{Cumulative distribution of the grain diameter for the
    glass beads (a) and the sand (b). Pictures of the granular
    materials are shown in inserts. The median size of the grains is
    about 280 $\mu$m for the glass beads and 330 $\mu$m for the sand.}
  \label{beads}
\end{figure}

Experiments have been performed for two different type of grains:
glass beads (Silibeads, Figure~\ref{beads}(a)) and sieved sand
(Sifraco, Figure~\ref{beads}(b)) with a similar range of diameters:
200-400 $\mu$m as can been seen on the cumulative size distributions
for the two materials in Figure~\ref{beads}. We observe in the insert
pictures in Figure~\ref{beads} that the glass beads have a lot of
defects and that the main difference between the two kind of grains is
that the sand is faceted whereas the glass beads are rather
spherical. The angle of avalanche for the glass beads is about
$29^{\circ}$ and for sand about $35.5^{\circ}$. The cumulative
  distributions show the presence of few particles in the range
  50-100~$\mu m$ for the two types of granular material. Since the
dynamics of precursors is in agreement with previously reported
experiments with larger
beads~\cite{nerone03,zaitsev08,gibiat09,kiesgen12}, we may think that
the details of the particle size distribution is not crucial.

Three different levels of compaction have been used. For the lowest
compaction samples, a grid (mesh size 4~mm) is placed at the
bottom of the empty box, the granular material is gently poured in the
box, and then the grid is slowly lifted through the sample. More
compacted configurations are obtained from this initial loose
preparation by tapping the filled box in order to increase the density
of the system. Grids of different mesh sizes and forms have
  been tested not to change the global phenomenology. The less
compacted samples have packing fraction around $C \simeq .58 \pm .02$,
the intermediate case (called ``normal'' in the following) has $C
\simeq .61 \pm .02$, and the highest packing fraction is close to $C
\simeq .64$.

Four different relative humidity conditions have been used. Ambient
humidity was between 25 \% and 45 \% (called ``normal'' condition in
the following). We obtain a humidity smaller than 20 \% by leaving the
granular sample in an oven overnight and then letting it cool down in
presence of desiccant. We obtain high humidity by leaving the granular
sample overnight in an enclosure in the presence of either saturated
salt water or pure water. The relative humidity in the enclosure is
controlled by the nature of the salt, and we then obtain samples
prepared at respectively 70 \% and above 80 \% humidity. For
  all those values of the relative humidity, the dynamics is unchanged
  when modifying the duration of the experiment. So we can neglect
relative humidity variations at the timescale of the experiment.

\section{Experimental results}\label{sec3}
\subsection{Main phenomenology}\label{sec3.1}

\begin{figure}[htbp]
  \centering \includegraphics[width=\columnwidth]{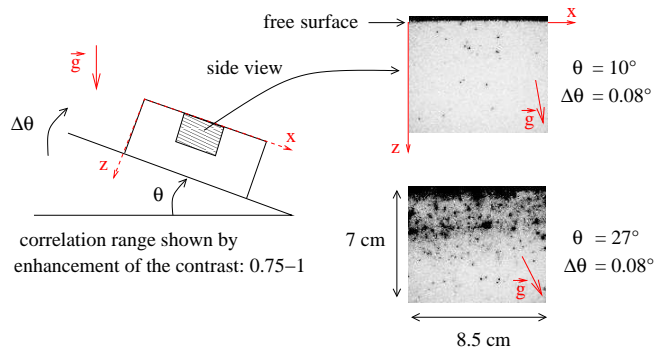}
  \caption{Schematic of the area used in the movie in Supplemental
    Material~\cite{movie}. Two examples of correlation maps extracted
    from the film, for angles 10$^{\circ}$ and 27$^{\circ}$, are shown
    and exhibit typical \emph{rearrangements}. The decorrelation maps
    of these images have been thresholded for $G_I$ smaller that 0.75
    to allow the rearrangements to be clearly identified. The granular
    material is glass beads, the packing fraction around .61 and
    hygrometry of 25 \%. The convention used for the axes are
    indicated on the figure.}
  \label{avalanche}
\end{figure}

A typical movie of the observed behavior in the so-called ``normal''
conditions (packing fraction around .61 and relative humidity of 25
\%) is given in Supplemental Material~\cite{movie}. The granular
material used in this movie is glass beads but the phenomenology is
very similar to sand. Each image is a map of the correlation between
two successive images and corresponds to the incremental deformation
during an angle variation of 0.08$^{\circ}$. The color scale is the
following: white correspond to $G_I=1$ (maximal correlation) and black
to $G_I=0$ (vanishing correlation). The value of the angle of
inclination of the board is indicated in degrees at the bottom-left of
the film. The free surface of the granular pile can clearly be
identified due to the fact that the light that does not come from the
sample is totally decorrelated. The size of the area seen in the film
is 7 cm $\times$ 8.5 cm from the upper middle of the box (see the
schematic in Figure~\ref{avalanche}). In order to evidence the small
rearrangements, the contrast of the maps shown on
Figure~\ref{avalanche} has been enhanced using a threshold value of
$G_I = 0.75$, so that all the correlations smaller than that value
appear in black.

Two distinct types of phenomenology can be observed. Firstly, spots of
different sizes appear well before any macroscopic event in the
sample. Such events are shown for example on Figure~\ref{avalanche}
for the 10$^{\circ}$ angle. Enhancement of the contrast shows that
such events appear as soon as the inclination process begins. These
spots are seemingly of the same nature as the \emph{rearrangements}
described previously in the literature~\cite{nerone03} and which were
observed at the free surface of the sample. Our observation tends to
show that those rearrangements happen in the whole bulk and are not
mere displacements of grains due to irregularities of the free surface
originating from the preparation of the sample, as has been postulated
in previous studies. The density and the intensity of those events
increase with the angle. The density seems also to depend on the depth
in the sample. Such localised events are reminiscent of the
\emph{spots} observed in a shear granular sample in~\cite{amon12}. The
fact that such spots have been observed at different borders in
several systems limited by different boundary conditions supports the
hypothesis that such localized events happen in the bulk of the
sample. A quantitative estimation of the energy released locally
compared to the overall work done during creep processes~\cite{amon12}
also supports that affirmation. The spots were identified
in~\cite{amon12} as the localized events of deformation introduced by
Argon~\cite{argon79} to describe the plasticity of amorphous
materials.

\begin{figure}[hbtp]
  \centering
  \includegraphics[width=\columnwidth]{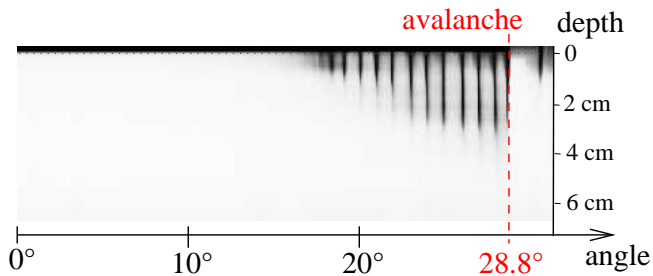}
  \caption{Spatiotemporal diagram obtained from the correlations map
    by spatially averaging each horizontal line of the images:
    $\langle G_I (\theta;\theta + \Delta \theta ; x ;z)\rangle_x$. The
    averaging procedure gives the mean decorrelation as a function of
    depth for each image. The granular material is glass beads, the
    packing fraction around .61 and relative humidity of 25
    \%. Correlations are calculated between images acquired at angles
    $\theta$ and $\theta+\Delta \theta$, with $\Delta
    \theta=0.08^{\circ}$.}
  \label{ST}
\end{figure}

Secondly, from an angle of about 15$^{\circ}$, large and almost regular
events begin to happen, which appear as successive large decorrelation
zones. These events correspond to the periodic \emph{precursors}
already described in the literature~\cite{nerone03}. Our experiment
makes it evident that these precursors involve a large part of the
bulk. These precursors begin to happen from about 15$^{\circ}$ and
occur nearly regularly, with an angular periodicity between
1$^{\circ}$ and 2$^{\circ}$. Quantitatively, these values correspond
well to the values that have been reported in the literature for other
sizes of grains and
containers~\cite{nerone03,zaitsev08,gibiat09,kiesgen12}. We can
observe that a precursor mobilizes a slice of the sample parallel to
the free surface. Such precursors are in fact micro-rupture events in
the bulk material. We also observe that the position of the
micro-rupture occurs at a depth that increases with each consecutive
event. Between two precursors, local rearrangements are also still
observed. Because of the approximative translational invariance of the
phenomenon along the optical plane, we obtain a spatio-temporal
representation by averaging the correlations at constant depth in each
image along horizontal lines ($x$-axis): $\langle G_I (\theta;\theta +
\Delta \theta ; x ;z)\rangle_x$. This gives a 1D representation of the
dependence of the correlation with depth for each image. By
juxtaposing these 1D lines for successive angles of inclination, we
obtain spatio-temporal graphs such as the one shown on
Figure~\ref{ST}. As the dependence between time and angle is linear in
a good approximation, the horizontal axis can be considered as
representing either time or angle. On such a representation the
periodic precursors are displayed distinctively while the
rearrangements contribute to an average decrease of the correlation,
giving a blurred or shaded aspect to the image where the activity in
terms of small rearrangements is high.

It is noticeable in the movie or in Figure~\ref{avalanche} at the
angle of $27^{\circ}$ that before and after a micro-rupture event
occurs, a large amount of activity can be observed at the depth where
the failure will take place. Larger spots of decorrelation are
observed in the zone that will fail compared to those that appear at a
larger depth. The decorrelation events seem to align and cluster at
the place where the rupture will take place, reminiscent of the
phenomenology already observed in a sheared granular
material~\cite{nguyen11,amon12} where regular micro-ruptures were also
observed before the final yield stress of the material is attained.

In the following section we give a more detailed study of that
global phenomenology which is very robust and resistant to changes of parameters
(grain shape, relative humidity, packing fraction). Firstly we will detail the
precursors events, then we will describe the rearrangements. Finally, we will study the changes of the phenomenology with
hygrometry and packing fraction.

\subsection{Precursors}\label{sec3.2}
\subsubsection{Periodicity}
\begin{figure}[htbp]
  \centering
  \includegraphics[width=\columnwidth]{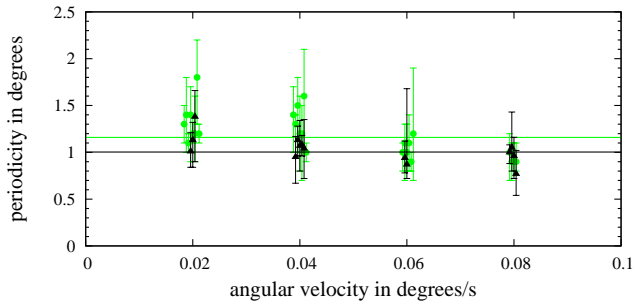}
  \caption{(Color online) Average period of the precursors with the
    tilt rate of the sample. The error bars give the extremal values
    of the period around the mean value during one experiment. Green
    (light gray) points are for glass beads, black triangles for
    sand. The horizontal lines give the average value of the period
    over all the runs for each type of material. All the experiments
    have been done with a similar preparation: packing fraction about
    0.61 and relative humidity about 30~\%}
  \label{period}
\end{figure}

We have studied experimentally the dependence of the periodicity with
respect to the tilt velocity. Figure~\ref{period} shows the average
period of the precursors in degrees for different rotation rates for
the same conditions of relative humidity and packing fraction
(so-called ``normal'' conditions: packing fraction about 0.61 and
relative humidity about 30 \%). Each point corresponds to the average
period between the precursors during one run. The value of the
period slightly decreases when the tilt rate increases, but overall
the phenomenon seems not to depend on the tilt rate. This mean that
the experiments can be considered as having been carried out in the
quasi-static limit. We also see in Figure~\ref{period} that the nature
of the material, glass beads (green (light gray) points) or sand
(black triangles), does not seem to modify significantly the overall
mean value of the period which is a little over 1$^{\circ}$ for both
materials. The error bars give the minimal and maximal values of the
period during each run. That dispersion around the mean value follows
a trend during one experiment: a significative increase of the period,
given by the size of the error bars, has been observed between the
first occurrence of the precursors around 15$^{\circ}$ and the
avalanche angle.

To conclude, the periodicity does not depend of the tilt rate in the
quasi-static limit for which the experiments have been done. It is
therefore justifiable to express that periodicity in terms of angles
and not in terms of time intervals. The period has been found to be
about 1$^{\circ}$ for the two types of granular materials. It
  has to be noted that similar values of the period have been obtained
  for much larger beads (2 to 3 mm), in containers of different sizes
  and materials~\cite{nerone03,zaitsev08,gibiat09,kiesgen12}.

\subsubsection{Size of the precursors}
\label{precursors}
An important question that remained unanswered in previous works was
whether the bulk is involved in the rearrangements and precursors
phenomena. As a matter of fact, most of the previous experiments have
been carried out using only free surface
observations~\cite{nerone03,kiesgen12}, so that rearrangements and
precursors could be supposed to be a mere free surface phenomenon. On
the other hand, acoustic measurements~\cite{zaitsev08,gibiat09} gave
rather indirect indications of the bulk mobilization. In our
experiment we are able to visualize directly how deep the phenomena go
inside the sample, even if the observation is confined to a
  thin layer near the wall of the container. We show that our
  results reinforce the hypothesis that a part of the bulk is indeed
  mobilized in a precursor event. The volume that is involved
  increases as the system comes closer to the avalanche angle.

\begin{figure}[htbp]
  \centering
  \includegraphics[width=3in]{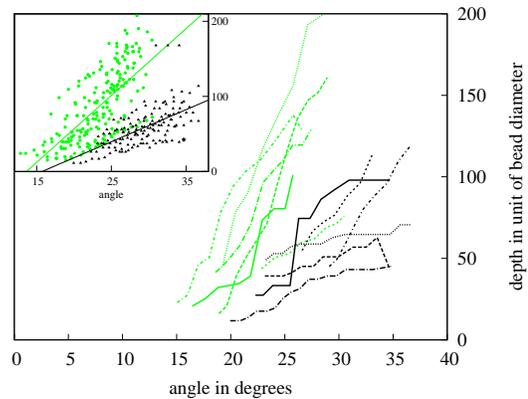}
  \caption{(Color online) Depth $z$, in units of grain diameter $d$,
    of the successive precursors plotted as a function of the angle of
    inclination of the granular sample $\theta$. All the experiments
    have been done for ``normal'' compaction ($\sim.61$) and relative
    humidity conditions ($\sim30-40$ \%) but different tilt
    rates. Main graph: examples of evolution for glass beads (green
    (light gray) lines) and sand (black lines) showing the roughly
    linear dependence of the depth with the angle. Insert: gathered
    results for a total of 25 experiments for glass beads (green
    (light gray) points) and 14 experiments for sand (black
    triangles). Lines are linear fits of those data: for glass beads,
    $z/d=9\times(\theta-14^{\circ})$, for sand: $z/d=4\times
    (\theta-16^{\circ})$.}
  \label{depth}
\end{figure}

The successive depths of precursors can be obtained from the
spatiotemporal graphs such as the one of Figure~\ref{ST}. The values
of the depth of the peaks, normalized by the diameter of the grains,
as a function of the angle at which they appear are shown on
Figure~\ref{depth}. The main part of the graph shows the evolution of
those depths for different typical experiments. We can observe that
the dependence of the size of the precursors with the inclination
angle is roughly linear, even if in some experiments brutal ``jumps''
can be observed (results displayed with solid lines). The green (light
gray) lines correspond to experiments with glass beads and the black
lines to the ones with sand. The slope of increase of depth with the
angle is generally larger for the glass beads than for the sand. Also,
the first precursor tends to appear sooner for glass beads than for
sand. The insert of Figure~\ref{depth} shows all the results from 25
experiments for glass beads and 14 for sand superimposed, giving an
idea of the dispersion of the results from one experiment to the
other. All the experiments have been done in the ``normal'' conditions
of relative humidity and packing fraction and for tilt rates between
0.02$^{\circ}$/s and 0.08$^{\circ}$/s. Linear fits using the whole set
of data for each type of granular material have been done and are
shown as solid lines in the insert. The slope of the line is about 9
beads diameter per degree for the glass beads and 4 for the sand. For
the two types of material, the intersection of the linear fit with the
abscissa axis is about 15$^{\circ}$, even if the first precursors
appear later in sand than in glass beads.

\begin{figure}[htbp]
  \centering
  \includegraphics[width=\columnwidth]{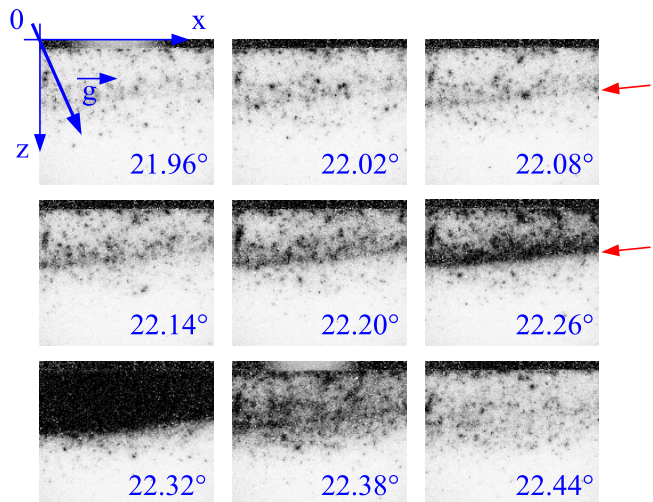}
  \caption{Detailed sequence describing a micro-slip event for an
    experiment with glass beads at ``normal'' hygrometry and packing
    fraction condition. The images shown are consecutive for angles
    around 22$^{\circ}$. The angle increment between the states used
    to calculate each correlation map is $0.06^{\circ}$. The red arrows show the localisation of the failure plane.}
  \label{rupture}
\end{figure}

Thus, precursor events involve a fraction of the material that
increases roughly linearly with the angle of inclination. That
increase is approximatively two times larger for the glass beads than
for the sand. The part of the sample that is mobilized is parallel to
the surface, such that the process seems to imply a rectangular block
of granular material in the upper part of the sample. A closer
observation of the movie in Supplemental Material~\cite{movie} shows
that the surface of separation between the mobilized zone and the
quiescent zone bends toward the bottom of the box, which could be an
effect of the side boundaries. A more detailed description of the
processes happening in the decorrelated area during a precursor event
is difficult giving the impossibility of separating the contributions
from deformation, plastic rearrangements and large scale displacement
of the block to the decorrelation. Nevertheless, some hints of the
process can be inferred from a sequence before a micro-rupture, as
shown in Figure~\ref{rupture}. We can see on this figure a ''rupture''
occurring at an angle of $22.32^{\circ}$. Yet the failure
  happens before as evidenced by a decorrelated zone mainly along a
  line (see arrow on fig.~\ref{rupture} \@ $22.26^{\circ}$). This line
  is indicative of a zone of shearing between two translating blocks.
As a matter of fact, a small solid translation does not modify the
optical paths relative to each other and the images stay
correlated. Nevertheless, the two blocks are not mechanically
identical: the activity in the upper section is clearly larger than in
the bottom. Finally, the micro-rupture event occurs. It is not
temporally resolved and appears as a large totally decorrelated
area. Please note that the total time of the sequence is $\sim~3$~s,
which is very large compared to the inertial time corresponding to a
displacement of, say, one bead diameter $\sqrt{d/g}~\simeq~5$~ms. So,
inertial effects do not seem to play a role in the development of the
micro-rupture. The phenomenology stays globally the same from one
experiment to the other, either for sand or glass beads: the density
of rearrangements increases in the area where the rupture will take
place, nevertheless a line of decorrelation at some depth of the
sample as the one distinctly seen at 22.26$^{\circ}$ on
Figure~\ref{rupture} is not always clearly seen.

We thus observe that the density of rearrangements and the large
periodic precursors events are not independent. Consequently, in the
following section we will focus on the quantitative study of the
rearrangements in the sample.

\subsection{Analysis of the rearrangements}\label{sec3.3}
\subsubsection{Density of activity in the material}
\label{activity}

\begin{figure}[htbp]
  \centering
  \includegraphics[width=\columnwidth]{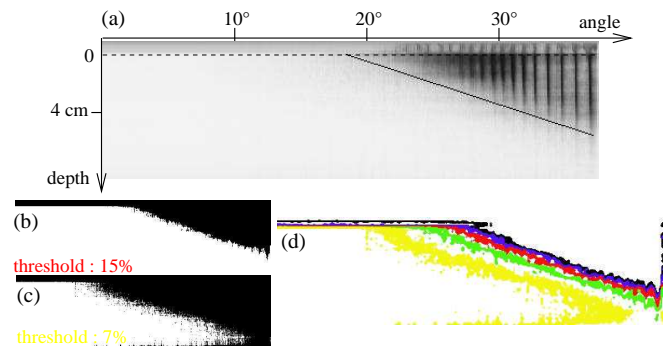}
  \caption{(Color online) (a) Spatio-temporal representation of the
    $x$-averaged correlation $G_I$ as a function of the depth and of
    the inclination angle. The material is sand in ``normal''
    conditions. The dashed line gives the mean slope of the fronts
    extracted from the binarized images (see main text). (b) and (c)
    Same data binarized with two different thresholds: (b) 0.15 and
    (c) 0.07. (d) Fronts extracted from the binarized
    spatio-temporal graphic for different values of the threshold,
    the gray scale is lighter when the threshold is smaller;
    yellow: 0.07, green: .11, red: .15, blue: .19, black: .23.}
  \label{front}
\end{figure}

To evidence the level of activity in the samples during the
inclination process, we first threshold the spatio-temporal images at
different levels of the correlation, giving an insight into the
distribution of the deformation in the depth of the sample during the
inclination. Figure~\ref{front} shows how we proceed. The original
spatiotemporal graph is shown on
Figure~\ref{front}(a). Figure~\ref{front}(b) and (c) are binarized
figures obtained from Figure~\ref{front}(a) for different values of
the threshold. For Figure~\ref{front}(b), all values of correlations
smaller (respectively greater) than 0.15 are black (resp. white). In
the case of Figure~\ref{front}(c), the procedure is the same with a
value of the threshold for the binarization of
0.07. We observe that during the
inclination process the level of deformation evolves as a front in the
sample: the depth at which a fixed level of deformation is reached in
the sample depends linearly on the angle of
inclination. Figure~\ref{front}(d) shows a superposition of the fronts
extracted from the binarized images for five different values of the
thresholds. We observe that the slope of the front remains
approximately constant. The solid line Figure~\ref{front}(a) is
obtained by averaging the slopes of the fronts from
Figure~\ref{front}(d). This shows that the growth of the precursors
events is identical to the overall repartition of the deformation in
the sample. Indeed, the average slope over several experiments, all
carried out in ``normal'' conditions of packing fraction and relative
humidity, is $13~d/^{\circ}$ for glass beads and $8~d/^{\circ}$ for
sand. Those slopes are of the same order and in the same ratio as the
values of the variation of the depth of the precursors with angle
($9~d/^{\circ}$ for glass beads and $4~d/^{\circ}$ for sand)
determined from data plotted on Figure~\ref{depth}.

\begin{figure}[htbp]
  \centering
  \includegraphics[width=\columnwidth]{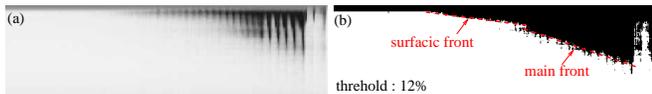}
  \caption{(a) Spatio-temporal representation of the x-averaged
    correlation $G_I$ as a function of the depth and of the
    inclination angle. The material is sand in ``normal''
    conditions. (b) Binarized image obtained from (a) for a threshold
    of 0.12.}
  \label{2pentes}
\end{figure}

In addition to this front which shows a linear relation between angle
and depth at a value of the deformation and which superimposed on the
growth of the precursors at large deformations, a lot of experiments
display another deformation front very close to the surface. Such a
phenomenon is certainly linked to a boundary effect in the vicinity of
the free surface. An example of such behavior is shown on
Figure~\ref{2pentes}. We observe in that figure that apart from the
main front that overlaps the precursors, a thin zone of deformation
breaks the slope of the front near the surface.

To conclude, the deformation in the sample appears as a front in a
spatio-temporal representation. That front appears very soon in the
inclination process and superimposed on the precursors peaks at large
deformation. The linear relation that connects the angle and the depth
at which the processes are observed seems to be of the same nature as
the linear increase of the size of the precursors with the angle of
inclination.

\subsubsection{Spots identification}

\begin{figure}[htbp]
  \centering
  \includegraphics[width=\columnwidth]{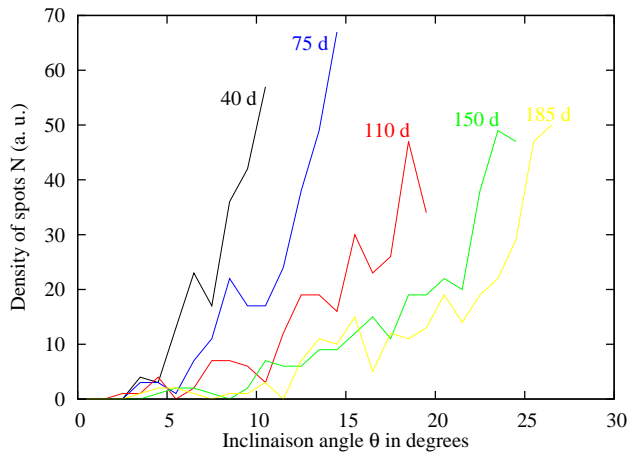}
  \caption{(Color online) Number of spots
      $N$ as a function of the angle of inclination $\theta$ at
      different depths in the sample. The average depth (in beads
      diameter unity $d$) at which the measurement has been made is
      indicated near each curve. The counting has been done
      integrating events during 1$^{\circ}$ inclination and over a
      slice of $\Delta z \simeq 37~d$ in depth.}
  \label{fluidity}
\end{figure}

Another way to quantify the activity in the material is to count the
number of \emph{spots}, \emph{i.e.} localized deformation in the
sample, and to look at its evolution during the tilt process. For
this, we threshold each image in a movie, and then we identify those
spots in each image. We then count the number of spots that have
appeared at a certain depth during an angle increment of
1$^{\circ}$. The number of spots $N$ occurring between $z-\Delta z/2$
and $z+\Delta z /2$, with $\Delta z$ the size of the slice, can be
measured for different depths $z$, as a function of the current angle
$\theta$. The result of that image analysis is given in
figure~\ref{fluidity} where the number of spots at different depths is
given as a function of the angle of inclination. Each line corresponds
to a different depth, from 40~$d$ under the surface to 185~$d$
deep. For a given depth the counting is stopped as a precursor
reaching that depth occur. We observe that at a fixed depth the number
of spots increases with the angle while at fixed angle it decreases
with the depth. At all the depths, the density of spots increase
strongly when approaching the precursor event at that depth. The final
values of the densities are of the same order at all the depth.

A further analysis of those data will be done in the
  discussion part. The first conclusions from the raw data are that
  the rearrangements in the sample begin as soon as the inclination
  process begins, with a decreasing density in the depth of the
  sample. The density of spots increases strongly before the occurrence
  of a precursor event at the corresponding depth.

\subsection{Influence of parameters}\label{sec3.4}

In this section, we will expose the consequences of the modification
of the humidity and packing fraction on the above observations.

\subsubsection{Influence of humidity}
\begin{figure}[htbp]
  \centering
  \includegraphics[width=\columnwidth]{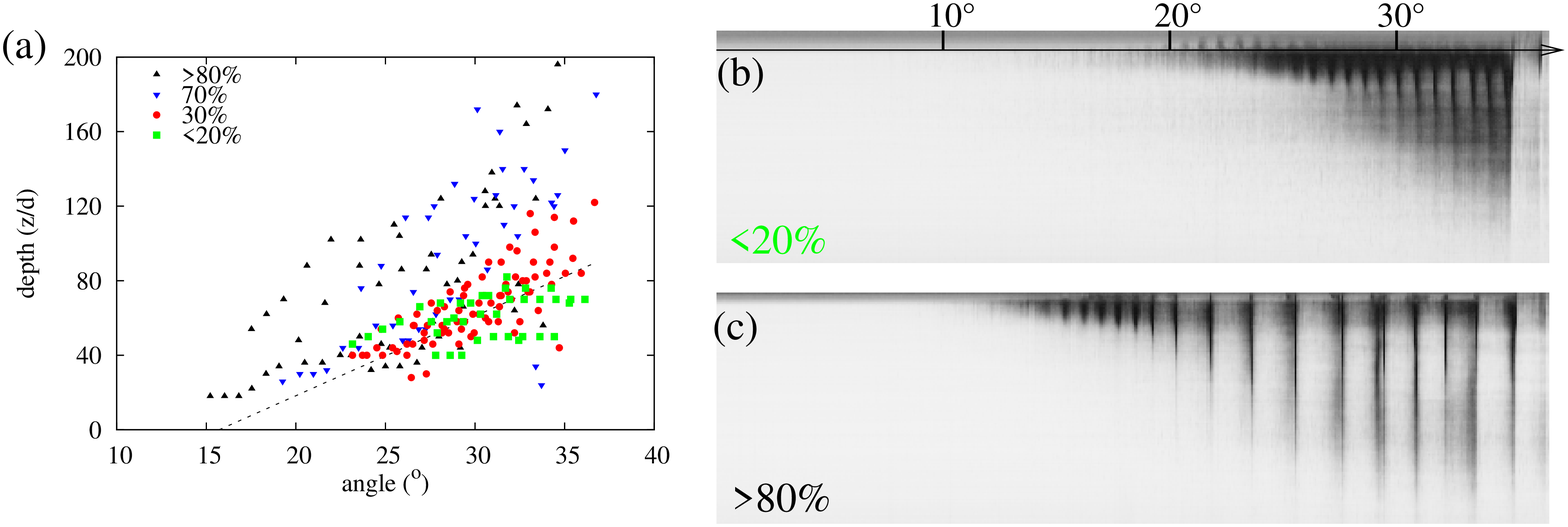}
  \caption{(Color online) (a) Depth of the successive precursors as a
    function of angle for sand samples at different humidity
    preparation. Higher values of humidity mobilize more grain layers
    and the precursors begin at smaller angles. (b) and (c):
    Spatio-temporal diagrams of sand specimens prepared at different
    humidity levels. (b) Dry case (hygrometry $<$~20~\%); (c) Humid
    case (hygrometry $>$~80~\%). The packing fraction preparation of
    the samples is the same.}
  \label{humidity}
\end{figure}

Figure~\ref{humidity} shows, for sand samples with different humidity
preparations, the size of the successive precursors as a function of
the angle of inclination. The experiments have been carried out with
sand prepared in the same manner as previously leading to a packing
fraction around $.61$ for the ``normal'' hygrometry conditions. The
dispersion of the results is rather large but it can be noticed that
the presence of humidity tends to increase the number of layers of
grains implied in the precursory process. This is in agreement with
the intuitive idea that the presence of humidity increase the cohesion
of the system and consequently that more grains are involved when a
rearrangement or a precursor occurs.

Other trends can also be noted: precursors appears sooner in a humid
system than in a dry one. In fact, precursors are better defined in
the humid case, as can be seen by comparing the spatio-temporal
diagram of an experiment with a dry sample (Fig.~\ref{humidity}(a))
and a humid sample (Fig.~\ref{humidity}(b)). In the dry case more
rearrangements take place giving a blurred appearance to the
diagram. The period of the precursors is also significantly smaller in
the dry case.

\begin{figure}[htbp]
  \centering
  \includegraphics[width=\columnwidth]{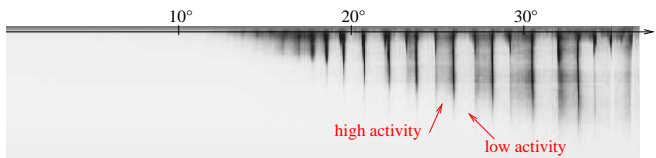}
  \caption{Spatio-temporal diagram of sand specimen prepared at high
    humidity level (above 70 \%): the precursors tends to appear by
    pairs showing a different
    rearrangement behavior periodically.}
  \label{2T}
\end{figure}

For humid cases, a curious pairing of precursors has been observed
several times, an example of which is shown in Figure~\ref{2T}. An
oscillation between two levels of activity can be observed, showing an
alternation between two different regimes after each precursors. We
also observed in some experiments an alternatation of two sizes for
the precursors, which explain for example the appearance in
Figure~\ref{humidity} of very small values of the precursors near the
avalanche in some humid cases: these values alternate with much higher
values.

\subsubsection{Influence of packing fraction}
When exploring different types of packing fraction preparation, we observed
that in the case of samples prepared in the loosest state, the
phenomenology is the same as in the ``normal'' conditions. In that
case the activity in the system is very high through the depth of the
pile and precursors are still observed. On the contrary, the
phenomenology is modified in the most compacted piles. In that case,
because of the significant increase of the avalanche angle for dense
materials, we were not able to reach the avalanche with our
experimental setup in most of the experiments.

\begin{figure}[htbp]
  \centering \includegraphics[width=\columnwidth]{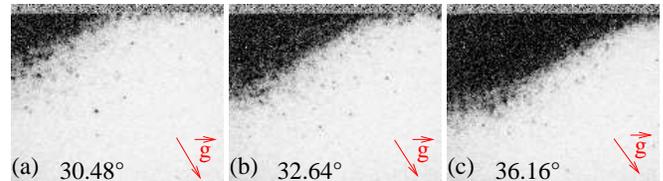}
  \caption{Maps of correlation at different angles for a sample of
    sand prepared in a compacted state at ``normal'' hygrometry
    conditions (32 \% relative humidity): (a) 30.48$^{\circ}$, (b)
    32.64$^{\circ}$ and (a) 32.16$^{\circ}$.}
  \label{compact}
\end{figure}

In the compacted experiments, local rearrangements during the
  inclination of the system are still observed but the activity in the
  system is much lower than in the less dense samples. On the
  contrary, large micro-failures are not always observed and when they
  are, they differ from the precursors events described
  previously. For about half of our experiments, no large events were
  observed at all. For the other half, large micro-slips event were
  observed with the main difference that the mobilized area has a
  clear angle with the surface of the sample instead of being parallel
  as was the case in the previously described micro-slips. Such a
  mobilized zone can be seen in Figures~\ref{compact}(a) to (d), where
  the angle of the prismatic mobilized region to the surface is about
  30$^{\circ}$, so that the surface of rupture is almost horizontal in
  the laboratory frame. The mobilized zone progresses in the system as
  the angle of inclination increases. This progression occurs by
  regular steps at a similar period to the precursors obtained in
  samples prepared in ``normal'' conditions.

To conclude this section concerning the influence of preparation over
the pre-avalanche phenomenology, we observe that the global picture of
the process, \emph{i.e.} the local rearrangements in the whole depth
of the system and periodic micro-ruptures appearing every few degrees
from an inclination of $15^{\circ}$, remains unchanged over a wide range of
conditions of preparation. The phenomenon is very robust to changes in
packing fraction and humidity conditions. Only the most compacted systems
behave differently, presenting an internal rupture that progresses in
the bulk of the system at an angle to the surface of the material.

\section{Discussion}\label{sec4}
In the preceding section, we have described the complex dynamic
behavior preceding the macroscopic failure of a granular material. The
dynamics begin very early in the tilting process with isolated
rearrangements events. Such events have already been reported, but
with our sensitive side view characterization, we are able to obtain
information about their spatial distribution. At the very early stage
of the tilting process the activity is essentially limited to very
superficial layers, and as the tilting process progresses, the
activity progresses into the bulk. A striking feature is the presence
of many micro-failure events in the bulk. Such events occur quite
regularly during the tilting process. We first discuss this effect in
the framework of Mohr-Coulomb failure criterium. After this, we will
discuss how our observations may be related to the plasticity of
granular material.

\subsection{Cohesion}\label{sec4.1}
We first discuss the localization of failure in a tilted granular
material as it may be deduced from Coulomb failure criteria. The
equation of static equilibrium for a granular material is:
\begin{eqnarray}
{\partial \sigma_{xx} \over \partial x}+{\partial \sigma_{xz} \over \partial z}&=&\rho g \sin \theta\\
{\partial \sigma_{zx} \over \partial x}+{\partial \sigma_{zz} \over \partial z}&=&\rho g \cos \theta
\end{eqnarray}
where $\sigma_{xz}=\sigma_{zx}$ is the shear stress, $\sigma_{xx}$
($\sigma_{zz}$) is the horizontal (vertical) normal stress, $\rho$ the
density, and $g$ the gravity. The orientations of the axes are defined
in Figure~\ref{avalanche}, and we neglect any shear stresses in the
transverse $y$ direction. Assuming that the stresses are uniform in
the $x$ direction, we obtain:

\begin{equation}
\sigma_{zz}=\rho g z \cos \theta,~~~~\sigma_{zx}=\rho g z \sin \theta \label{eqStresses}
\end{equation}

The Coulomb criterion postulates that a granular material is stable with respect to failure if~\cite{nedderman}:
\begin{equation}
\vert \tau \vert \le \mu \sigma +c_h, \label{eqCoulomb}
\end{equation}
for any plane inside the material. In Equation~(\ref{eqCoulomb}),
$\tau$ is the shear stress, $\sigma$ the normal stress, $\mu$ the
internal friction coefficient, and $c_h$ the cohesion. Applying the
Coulomb criteria to Equation~(\ref{eqStresses}), we find that the
planes of failure are parallel to $x$ axis, and the Coulomb criteria
is now:

\begin{equation}
\tan \theta \le \mu +{c_h \over \rho g z \cos \theta}, \label{eqCoulomb2}
\end{equation}

First, in the absence of cohesion $c_h=0$, the depth of failure plane
is not determined. With cohesion, the position of the failure plane is
determined. In the case of constant cohesion, failure must occur at
the bottom of the sample~\cite{halsey98,restagno04}, where ${c_h /\rho
  g z \cos \theta}$ is the lowest. That result is in contradiction
with the observation that the first micro-failure appears at a smaller
depth than the following ones. The dependence of $\theta$ on the
values of $z$ for which failure occurs in a Mohr-Coulomb model is
opposite to the dependence experimentally observed. Moreover, such a
model gives only one criteria of rupture and is unable to predict
successive micro-failures in the material. In any case, the Coulomb
failure criterium is unable to explain the occurrence of the observed
micro-failures into the bulk of the sample.

\subsection{Plastic deformation}\label{sec4.2}

The Coulomb failure discussed in the preceding section does not take
into account any plastic deformations before rupture in the granular
material. It is however well known that granular materials may yield
before being subjected to failure. The critical state theory of
soil~\cite{schofield} discussed how the plastic deformation occurs in
a material depending on the initial state of the material. Starting
from an initially loose granular material, application of a shear
stress initially produces an elastic deformation. After this, yield
begins with plastic deformations. During this yielding, the material
compacts slowly. When the shear stress exceeds a threshold (generally
a fraction of the confining pressure), macroscopic failure occurs in
the material. If the experiment begins with an initially dense
granular material, the deformation is elastic until a maximum value of
the stress is reached. When the stress exceeds this value, failure
occurs. No plastic deformation happens before that rupture and
dilatancy occurs only after the failure. The fact that rearrangements
and precursors are observed in our experiments only for loose or
moderately dense granular samples (and not to densely prepared
granular materials) seems to show that the observed deformations in
our experiments correspond to yield before failure.  We observed in a
previous work, for a different experimental geometry,
that the same kind of rearrangements may be associated to plastic
dissipation into the material~\cite{amon12}.

\begin{figure}[htbp]
  \centering
  \includegraphics[width=\columnwidth]{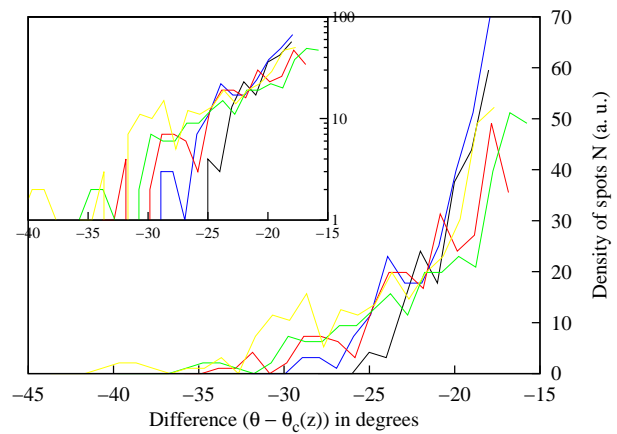}
  \caption{(Color online) Number of spots $N$ as a function of the
    difference between actual angle and failure angle
    $\theta_c(z)$. Every curve corresponds to a different depth: black
    $z=40d$, blue $75d$, red $110d$, green $150d$, yellow $185d$.
    Inset: same curves in logarithmic scale.}
  \label{fig14}
\end{figure}

The link between the plastic deformations and the failure of the
material may be highlighted by considering the amount of plastic
deformation as a function of the difference between the applied stress
and the stress at which failure occurs. We have noticed (see
sec.\ref{activity} and fig.\ref{front}) that, at a given tilt angle
(i.e. at a given applied stress), the density of the rearrangements
decreases with the depth. This dependence of the density of the
rearrangements on the depth of the material most certainly originates
from a gradient of properties, either due to a depth dependent
confining pressure and/or a gradient of packing fraction during the
preparation of the sample. Such packing fraction and/or pressure
dependence of the yield process is observed in many soil mechanics
studies, but at confining pressures noticeably larger than the
pressure in our experiment. The tilt angle at which failure occurs
into the material (i.e. the failure stress) also increases with the
depth (see sec.\ref{precursors} and fig.\ref{depth}). From data of
fig.\ref{depth}, we defined $\theta_c(z)$ as the $z$-dependent value
of the tilt angle at which failure occurs. we plot on Fig.~\ref{fig14}
the number of events as a function of $\theta - \theta_c(z)$. The data
obtained for different depths, which were scattered on
fig.\ref{fluidity}) then collapse on the same curve. This shows that
those plastic events occur with a density which is given by the
difference between the failure and applied stresses, or equivalently
that failures happen after a certain number of rearrangements have
occurred.

It has been shown in~\cite{amon12} that the rate at which localised plastic events occur may be identified with the {\it so-called fluidity} of the material, which is the local rate of stress relaxation. The concept of fluidity has been introduced in order to explain the rheological properties of soft glassy material~\cite{derec2001}. The use of this concept in order to explain rheological properties of granular material has also been recently proposed~\cite{kamrin2012,chaudhuri2012}. The relation between isolated plastic events and failure may be qualitatively understood. Indeed, since
rearrangements may be seen as plastic reorganizations, we expect that
each event redistributes some stress, accordingly to an elastic stress
propagator~\cite{picard2005}. This additional stress may also trigger
other plastic events in their neighborhood. Because of these
processes, shear bands may form in the material~\cite{martens2012}. It
follows naturally from this kind of scenario that failure occurs when
a certain amount of activity in the material is reached. This is
precisely what is observed: rearrangements precede micro-failure,
whatever is the depth.

\subsection{Periodicity of precursors}\label{sec4.3}

Another striking observation is a rough periodicity in tilt angle of
the precursor events. As we already shown (see sec.\ref{precursors}
and Fig.~\ref{rupture}), those precursors events are indeed failure in
the bulk material.

\begin{figure}[htbp]
  \centering
  \includegraphics[width=\columnwidth]{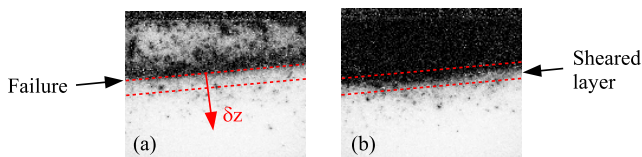}
  \caption{(Color online) Details of a slip event. (a) is the
    correlation image of a microslip event showing the failure
    plane. (b) is the following correlation image. The zone between
    the dashed red lines is decorrelated although below the failure
    plane, and may be interpreted as a layer of sheared granular
    material.}
  \label{fig15}
\end{figure}

We have plotted on Fig.~\ref{fig15} a detail of the failure already
described on Fig.~\ref{rupture}. We see on Fig.~\ref{fig15}.a the
localisation of the failure plane. One image latter
(Fig.~\ref{fig15}.b) we observe that the decorrelated zone extends
deeper into the material. So we have between the two lines of
fig.\ref{fig15}.b a layer of material which has been deformed and
which is below the failure plane. We can interpret this layer as a
zone of granular material which has been sheared by the motion of the
upper part. This observation is in agreement with the literature about
creep motion in granular flow. It is indeed well known that when a
granular layer is sheared, the deformation extends into the bulk. The
deformation decays exponentially into the bulk, with a characteristic
length $\xi$ of the order of few grain
diameters~\cite{komatsu2000,richard2008,crassous2008}. The same decay
in velocity is also observed during non stationary granular
avalanches~\cite{courrech2005}. So the effect of a micro-failure event
may be to produce a deformation which extends into the depth. This
interpretation of this decorrelated layer is supported by the
following quantitative analysis. Calling $\Delta$ the displacement of
the upper block, we may
expect~\cite{komatsu2000,richard2008,crassous2008,courrech2005} a
displacement which decays as $\simeq \Delta \exp{(-\delta z/\xi)}$,
with $\delta z$ the distance below the
  failure plane (see fig.~\ref{fig15}). The deformation
is then $\simeq (\Delta/\xi) \exp{(-\delta z/\xi)}$. With $\Delta \sim
\xi \simeq d$, we have a deformation of order $10^{-6}$ (the limit of
detection of our light scattering setup) for $\delta z \simeq ln(10^6)
\times \xi \simeq 15 d$. This is in rough agreement with the
separation between the two lines of fig.\ref{fig15}.b which is $\simeq
15 d$. This agreement supports the hypothesis that a layer of material
is sheared below the failure plane.

The effect of this shear deformation may be to erase all the possible
site for future micro-failure events because of the reconfiguration of
the sample in that zone. The next micro-failure site should then be
located slightly deeper into the bulk. The typical depth depth of this
sheared layer slip being $15~d$, we may expect from the data of
section \ref{precursors} (variation of precursor depths with failure
angle $\simeq 9~d/deg$ for glass beads) that the angular period
between successive micro-failure is of the order of $\simeq 15/9
\approx 1.6~deg$, a value which is close to the period that we
measured (see figure~\ref{period}).

\section{Conclusion}\label{sec5}
Our experiments on quasi-static tilting of granular materials show, in
agreement with numerous previously published studies, that
reorganizations occur before the avalanche takes place. The main
advance of this study is to show that such rearrangements are
organized spatially in a complex manner. For loose or moderately dense
granular systems, we observe isolated reorganizations at low
inclination, with a density decreasing slowly with the depth. As
inclination increases, reorganizations occur under the form of
micro-failure planes in the bulk, which are localized at increasing
depths. The different localizations of the failure planes imply that
the underlying physics is different from the stick-slip already
observed in other granular experiments~\cite{nasuno98}, where the slip
plane remains the same. The micro-failures seem to occur when a given
level of accumulated plastic deformation is attained. This holds at
every depth. The case of dense granular systems seems to obey to a
different underlying physics.

Those observations may be partially and qualitatively understood in the framework of
yielding properties of granular material (e.g. Granta-Gravel material
in~\cite{schofield}) before failure. At densities below the critical
one, the granular material yields progressively before the failure of
the material. This yielding is continuous, and associated to a
compaction of the material. We may then interpret the observed plastic
deformation as experimental manifestation of this yielding. Important
deformations are then observed in the form of micro-failures, which,
in this framework, form plastic flow in the material. The periodicity
of such plastic flows may be understood if we consider collective
granular flow: micro-failures involving granular matter in the bulk
with a typical extension length, and the next micro-failure may occur
in the next un-deformed zone. This explanation is evidently very partial and only qualitative. We do
not expect that a granular model such as Granta-Gravel will apply
completely or qualitatively, and it must be understood only as a
starting point for future investigations.

A puzzling point is that, to
our knowledge, such collective and regular precursors motions have not
been reported in numerical simulations of tilted granular material. It
follows from our experimental results that those collective events may
only be observed with moderately dense and large enough numerical
granular packing.

Another interesting point is that those micro-failures do not occur at
the beginning of the tilt process. There is some minimum angle
(i.e. some minimal shear stress) below which no micro-failures
occur. To our knowledge, the existence of two finite critical shear
stresses (one for micro-failure and one for the macroscopic failure)
is not discussed in the literature. A complete understanding of the
observation described in this paper appears as a challenging task.

\section{Acknowledgements}
We acknowledge the financial supports ANR ``STABINGRAM''
No. 2010-BLAN-0927-01. We thank Eric Cl\'ement, Lyd\' eric Bocquet,
Patrick Richard, Mickael Duranteau and Yves Le Gonidec for scientific
discussions, Patrick Chasle for help with the image acquisition and
Alain Faisant for the conception of the tilting optical board.

\bibliography{precursors}{}

\begin{thebibliography}{10}

\bibitem{jaeger89}
H.~M. Jaeger, C.~Liu, and S.~R. Nagel.
\newblock Relaxation at the angle of repose.
\newblock {\em Phys. Rev. Lett.}, 62:40--43, Jan 1989.

\bibitem{bretz92}
M.~Bretz, J.~B. Cunningham, P.~L. Kurczynski, and F.~Nori.
\newblock Imaging of avalanches in granular materials.
\newblock {\em Phys. Rev. Lett.}, 69:2431--2434, Oct 1992.

\bibitem{nerone03}
N.~Nerone, M.~A. Aguirre, A.~Calvo, D.~Bideau, and I.~Ippolito.
\newblock Instabilities in slowly driven granular packing.
\newblock {\em Phys. Rev. E}, 67:011302, Jan 2003.

\bibitem{nagel92}
S.~R. Nagel.
\newblock Instabilities in a sandpile.
\newblock {\em Rev. Mod. Phys.}, 64:321--325, Jan 1992.

\bibitem{fischer08}
R.~Fischer, P.~Gondret, B.~Perrin, and M.~Rabaud.
\newblock Dynamics of dry granular avalanches.
\newblock {\em Phys. Rev. E}, 78:021302, Aug 2008.

\bibitem{bak87}
P.~Bak, C.~Tang, and K.~Wiesenfeld.
\newblock Self-organized criticality: An explanation of the 1/ \textit{f}
  noise.
\newblock {\em Phys. Rev. Lett.}, 59:381--384, Jul 1987.

\bibitem{zaitsev08}
{Zaitsev, V. Yu.}, {Richard, P.}, {Delannay, R.}, {Tournat, V.}, and {Gusev, V.
  E.}
\newblock Pre-avalanche structural rearrangements in the bulk of granular
  medium: Experimental evidence.
\newblock {\em EPL}, 83(6):64003, 2008.

\bibitem{gibiat09}
V.~{Gibiat}, E.~{Plaza}, and P.~{De Guibert}.
\newblock {Acoustic emission before avalanches in granular media}.
\newblock {\em ArXiv e-prints}, June 2009.

\bibitem{kiesgen12}
S.~Kiesgen de~Richter, G.~Le Caër, and R.~Delannay.
\newblock Dynamics of rearrangements during inclination of granular packings:
  the avalanche precursor regime.
\newblock {\em Journal of Statistical Mechanics: Theory and Experiment},
  2012(04):P04013, 2012.

\bibitem{staron02}
L.~Staron, J.-P. Vilotte, and F.~Radjai.
\newblock Preavalanche instabilities in a granular pile.
\newblock {\em Phys. Rev. Lett.}, 89:204302, Oct 2002.

\bibitem{staron06}
L.~Staron, F.~Radjai, and J.-P. Vilotte.
\newblock Granular micro-structure and avalanche precursors.
\newblock {\em J. Stat. Mech.}, 2006:P07014, July 2006.

\bibitem{duran}
Duran J.
\newblock {\em Sands, powders, and grains}.
\newblock Partially ordered systems. Springer, New York, 2000.

\bibitem{carlson94}
J.~M. Carlson, J.~S. Langer, and B.~E. Shaw.
\newblock Dynamics of earthquake faults.
\newblock {\em Rev. Mod. Phys.}, 66:657--670, Apr 1994.

\bibitem{nasuno97}
S.~Nasuno, A.~Kudrolli, and J.~P. Gollub.
\newblock Friction in granular layers: Hysteresis and precursors.
\newblock {\em Phys. Rev. Lett.}, 79:949--952, Aug 1997.

\bibitem{nasuno98}
S.~Nasuno, A.~Kudrolli, A.~Bak, and J.~P. Gollub.
\newblock Time-resolved studies of stick-slip friction in sheared granular
  layers.
\newblock {\em Phys. Rev. E}, 58:2161--2171, Aug 1998.

\bibitem{ramos06}
O.~Ramos, E.~Altshuler, and K.~J. M\aa{}l\o{}y.
\newblock Quasiperiodic events in an earthquake model.
\newblock {\em Phys. Rev. Lett.}, 96:098501, Mar 2006.

\bibitem{rubinstein07}
S.~M. Rubinstein, G.~Cohen, and J.~Fineberg.
\newblock Dynamics of precursors to frictional sliding.
\newblock {\em Phys. Rev. Lett.}, 98:226103, Jun 2007.

\bibitem{rubinstein09}
S.~M. Rubinstein, G.~Cohen, and J.~Fineberg.
\newblock Visualizing stick–slip: experimental observations of processes
  governing the nucleation of frictional sliding.
\newblock {\em Journal of Physics D: Applied Physics}, 42(21):214016, 2009.

\bibitem{ben-David10}
O.~Ben-David, G.~Cohen, and J.~Fineberg.
\newblock The dynamics of the onset of frictional slip.
\newblock {\em Science}, 330(6001):211--214, 2010.

\bibitem{maegawa10}
S.~Maegawa, A.~Suzuki, and K.~Nakano.
\newblock Precursors of global slip in a longitudinal line contact under
  non-uniform normal loading.
\newblock {\em Tribology Letters}, 38:313--323, 2010.
\newblock 10.1007/s11249-010-9611-7.

\bibitem{schreibert10}
J.~Scheibert and D.~K. Dysthe.
\newblock Role of friction-induced torque in stick-slip motion.
\newblock {\em EPL (Europhysics Letters)}, 92(5):54001, 2010.

\bibitem{tromborg11}
J.~Tr\o{}mborg, J.~Scheibert, D.~S. Amundsen, K.~Th\o{}gersen, and
  A.~Malthe-S\o{}renssen.
\newblock Transition from static to kinetic friction: Insights from a 2d model.
\newblock {\em Phys. Rev. Lett.}, 107:074301, Aug 2011.

\bibitem{braun09}
O.~M. Braun, I.~Barel, and M.~Urbakh.
\newblock Dynamics of transition from static to kinetic friction.
\newblock {\em Phys. Rev. Lett.}, 103:194301, Nov 2009.

\bibitem{rubinstein11}
S.~Rubinstein, I.~Barel, Z.~Reches, O.~Braun, M.~Urbakh, and J.~Fineberg.
\newblock Slip sequences in laboratory experiments resulting from inhomogeneous
  shear as analogs of earthquakes associated with a fault edge.
\newblock {\em Pure and Applied Geophysics}, 168:2151--2166, 2011.
\newblock 10.1007/s00024-010-0239-1.

\bibitem{klaumunzer11}
D.~Klaum\"unzer, A.~Lazarev, R.~Maa\ss{}, F.~H. Dalla~Torre, A.~Vinogradov, and
  J.~F. L\"offler.
\newblock Probing shear-band initiation in metallic glasses.
\newblock {\em Phys. Rev. Lett.}, 107:185502, Oct 2011.

\bibitem{nguyen11}
V.~B. Nguyen, T.~Darnige, A.~Bruand, and E.~Clement.
\newblock Creep and fluidity of a real granular packing near jamming.
\newblock {\em Phys. Rev. Lett.}, 107:138303, Sep 2011.

\bibitem{amon12}
A.~Amon, V.~B. Nguyen, A.~Bruand, J.~Crassous, and E.~Cl\'ement.
\newblock Hot spots in an athermal system.
\newblock {\em Phys. Rev. Lett.}, 108:135502, Mar 2012.

\bibitem{erpelding08}
M.~Erpelding, A.~Amon, and J.~Crassous.
\newblock Diffusive wave spectroscopy applied to the spatially resolved
  deformation of a solid.
\newblock {\em Phys. Rev. E}, 78:046104, Oct 2008.

\bibitem{erpelding10}
M.~Erpelding, A.~Amon, and J.~Crassous.
\newblock Mechanical response of granular media: New insights from
  diffusing-wave spectroscopy.
\newblock {\em EPL (Europhysics Letters)}, 91(1):18002, 2010.

\bibitem{crassous07}
J.~Crassous.
\newblock Diffusive wave spectroscopy of a random close packing of spheres.
\newblock {\em Eur. Phys. J. E}, 23:145--152, 2007.

\bibitem{djaoui2005}
Linda Djaoui and J\'er\^ome Crassous.
\newblock Probing creep motion in granular materials with light scattering.
\newblock {\em Granular Matter}, 7:185--190, 2005.
\newblock 10.1007/s10035-005-0210-5.

\bibitem{movie}
see movie at url.

\bibitem{argon79}
Argon~A. S.
\newblock Plastic deformation in metallic glasses.
\newblock {\em Acta Metall.}, 27:47--58, Jan 1979.

\bibitem{nedderman}
Nedderman R.M.
\newblock {\em Statics and Kinematics of Granular Materials}.
\newblock Cambridge University Press, 1992.

\bibitem{halsey98}
T.~C. Halsey and A.~J. Levine.
\newblock How sandcastles fall.
\newblock {\em Phys. Rev. Lett.}, 80:3141--3144, Apr 1998.

\bibitem{restagno04}
F.~Restagno, L.~Bocquet, and E.~Charlaix.
\newblock Where does a cohesive granular heap break?
\newblock {\em The European Physical Journal E: Soft Matter and Biological
  Physics}, 14:177--183, 2004.
\newblock 10.1140/epje/i2004-10013-5.

\bibitem{schofield}
C.~P. Schofield, A. N.;~Wroth.
\newblock {\em Critical State Soil Mechanics}.
\newblock McGraw-Hill, 1968.

\bibitem{derec2001}
C.~Derec, A.~Ajdari, and F.~Lequeux.
\newblock Rheology and aging: A simple approach.
\newblock {\em The European Physical Journal E}, 4:355--361, 2001.

\bibitem{kamrin2012}
Ken Kamrin and Georg Koval.
\newblock Nonlocal constitutive relation for steady granular flow.
\newblock {\em Phys. Rev. Lett.}, 108:178301, Apr 2012.

\bibitem{chaudhuri2012}
Pinaki Chaudhuri, Vincent Mansard, Annie Colin, and Lyderic Bocquet.
\newblock Dynamical flow arrest in confined gravity driven flows of soft jammed
  particles.
\newblock {\em Phys. Rev. Lett.}, 109:036001, Jul 2012.

\bibitem{picard2005}
Guillemette Picard, Armand Ajdari, Fran\ifmmode \mbox{\c{c}}\else~\c{c}\fi{}ois
  Lequeux, and Lyd\'eric Bocquet.
\newblock Slow flows of yield stress fluids: Complex spatiotemporal behavior
  within a simple elastoplastic model.
\newblock {\em Phys. Rev. E}, 71:010501, Jan 2005.

\bibitem{martens2012}
Kirsten Martens, Lyderic Bocquet, and Jean-Louis Barrat.
\newblock Spontaneous formation of permanent shear bands in a mesoscopic model
  of flowing disordered matter.
\newblock {\em Soft Matter}, 8:4197--4205, 2012.

\bibitem{komatsu2000}
Teruhisa~S. Komatsu, Shio Inagaki, Naoko Nakagawa, and Satoru Nasuno.
\newblock Creep motion in a granular pile exhibiting steady surface flow.
\newblock {\em Phys. Rev. Lett.}, 86:1757--1760, Feb 2001.

\bibitem{richard2008}
P.~Richard, A.~Valance, J.-F. M\'etayer, P.~Sanchez, J.~Crassous, M.~Louge, and
  R.~Delannay.
\newblock Rheology of confined granular flows: Scale invariance, glass
  transition, and friction weakening.
\newblock {\em Phys. Rev. Lett.}, 101:248002, Dec 2008.

\bibitem{crassous2008}
J\'er\^ome Crassous, Jean-Fran\c cois Metayer, Patrick Richard, and Claude
  Laroche.
\newblock Experimental study of a creeping granular flow at very low velocity.
\newblock {\em Journal of Statistical Mechanics: Theory and Experiment},
  2008(03):P03009, 2008.

\bibitem{courrech2005}
Sylvain Courrech~du Pont, Rapha\"el Fischer, Philippe Gondret, Bernard Perrin,
  and Marc Rabaud.
\newblock Instantaneous velocity profiles during granular avalanches.
\newblock {\em Phys. Rev. Lett.}, 94:048003, Feb 2005.

\end{thebibliography}
\bibliographystyle{unsrt}

\end{document}